\documentclass[%
 reprint,
 amsmath,amssymb,
 aps,
 prl,
showkeys]{revtex4-2}

\usepackage{multirow}
\usepackage{siunitx}
\usepackage{graphicx}% Include figure files
\usepackage{dcolumn}% Align table columns on decimal point
\usepackage{bm}% bold math
\begin{document}

\preprint{APS/123-QED}

\title{Turbulence-based parametrization of animal flight}% Force line breaks with \\

\author{Ariane Gayout}
 \email{a.m.m.gayout@rug.nl}
\affiliation{%
 Groningen Institute for Evolutionary Life Sciences, University of Groningen, The Netherlands
}%


\author{Eize J. Stamhuis}
 \affiliation{
 Biomimetics group, Energy and Sustainability Research Institute Groningen, University of Groningen, The Netherlands
}

\author{Casper J. van der Kooi}%
\affiliation{%
Groningen Institute for Evolutionary Life Sciences, University of Groningen, The Netherlands
}%


\date{\today}% It is always \today, today,

\begin{abstract}
Animals capable of powered flight range in wingspan from a few hundred microns to a few meters. The inertial turbulence to which these animals are exposed features vortices ranging from a few hundred micrometers to hundreds of kilometers in size. Yet, the impact of ambient turbulence on animal flight is virtually uncharted and most studies on animal flight are conducted in still air or under laminar conditions. Here, we propose a novel parameterization that links animal flight with turbulence, through a proxy for the energy injected into the atmosphere, $E_{\rm sp}=b^3 f^2$, with $f$ the animal’s flapping frequency and $b$ the wingspan. We model this parameter using a scaling relation in the shape of a power law $E_{\rm sp} \propto k^\alpha$, with $k=1/b$ the wavenumber corresponding to the animal inverse wingspan. Literature provides four theoretical predictions on the exponent $\alpha$: two connected to aerodynamic and energetic aspects of flight, $\alpha_{\rm aero}=-2$ and $\alpha_{\rm power}=-5/3$, and two linked to physiological limits. Drawing from experimental data of over 400 species spanning 13 insect orders and two vertebrate classes, we recover $\alpha_{\rm power}=-5/3$ as the best scaling relation across the animal kingdom. Grouping per animal clade however reveals a secondary power law with $\alpha=-5/2$ exponent for invertebrate orders, with a family-dependent coefficient. This new scaling relation suggests a yet-unknown universal physical mechanism in insect flight, likely depending on wing morphology and mechanical properties. 
\end{abstract}

%Use showkeys class option if keyword
\maketitle

%\tableofcontents

Animal locomotion is usually studied from the animal perspective, focusing on its behavior and biomechanics. Yet, integrating the physical properties of the surrounding medium in locomotion models is crucial to better understand how environmental constraints influence animal motion and may act as a selective pressure. For instance, the granular properties of the ground shape the locomotor responses of insects, clams and snakes \cite{Pineirua2023, Hosoi2015}, whereas oceanic turbulence is known to impact plankton swimming on several levels \cite{Franks2024}. In the case of aerial locomotion, atmospheric properties, like atmospheric pressure or turbulence, impact significantly the flight performance of animals \cite{Sotavalta1947, Dudley1995, Combes2009, Engels2016, Engels2019}, but the integration of these parameters into animal flight models is still anecdotic \cite{Laurent2021, Shepard2025}. Incorporating the fluid surrounding the animal during its movement has resulted in the establishing of dimensionless scaling relations coupling the Reynolds and Strouhal numbers \cite{Taylor2003, Gazzola2014}.

For flight, many scaling relations have been proposed, linking the different morphological parameters and flight kinematics to the force production \cite{Jensen2024, Liu2025, Pohly2025}. These models have so far rarely included air properties (both extensive and intensive) and surprisingly, little consensus has been reached on the scaling exponents \cite{Darveau2024}. Promoted by large experimental collections started in the late 19th century \cite{Pettigrew1874, Marey1890, Magnan1934, Sotavalta1947, Greenewalt1962, Ellington1991, Tercel2018}, several frameworks have been proposed to explain animal flight kinematics, primarily grounded in theory based on aerodynamics \cite{Magnan1934, Rayner1979, Ellington1996}. Physiological concepts, like body/wing allometry and scaling effects, were further included in the proposed models, driven by mass variations across individuals and species \cite{Sotavalta1947, Schmidt-Nielsen1975}. Now, recent attempts at developing a framework that unifies flight kinematics across animal species have focused on universal scaling relations across locomotion types, including swimming and terrestrial locomotion to flight \cite{Bejan2006, Bale2014, Jensen2024, Liu2025}. These approaches mostly focus on the animal itself and the energetic cost of locomotion. Yet, scaling laws not only give information on the physiological aspects of locomotion but also help identify the key physical mechanisms at play, provided a physics-informed model is associated to the scaling exponent. In the context of animal flight, the proposed scaling exponents are often empirical  \cite{Darveau2024} and may therefore lack an associated physical model, despite our progressing understanding on flight physical mechanisms.

\begin{figure*}
    \centering
    \includegraphics[width=0.80\linewidth]{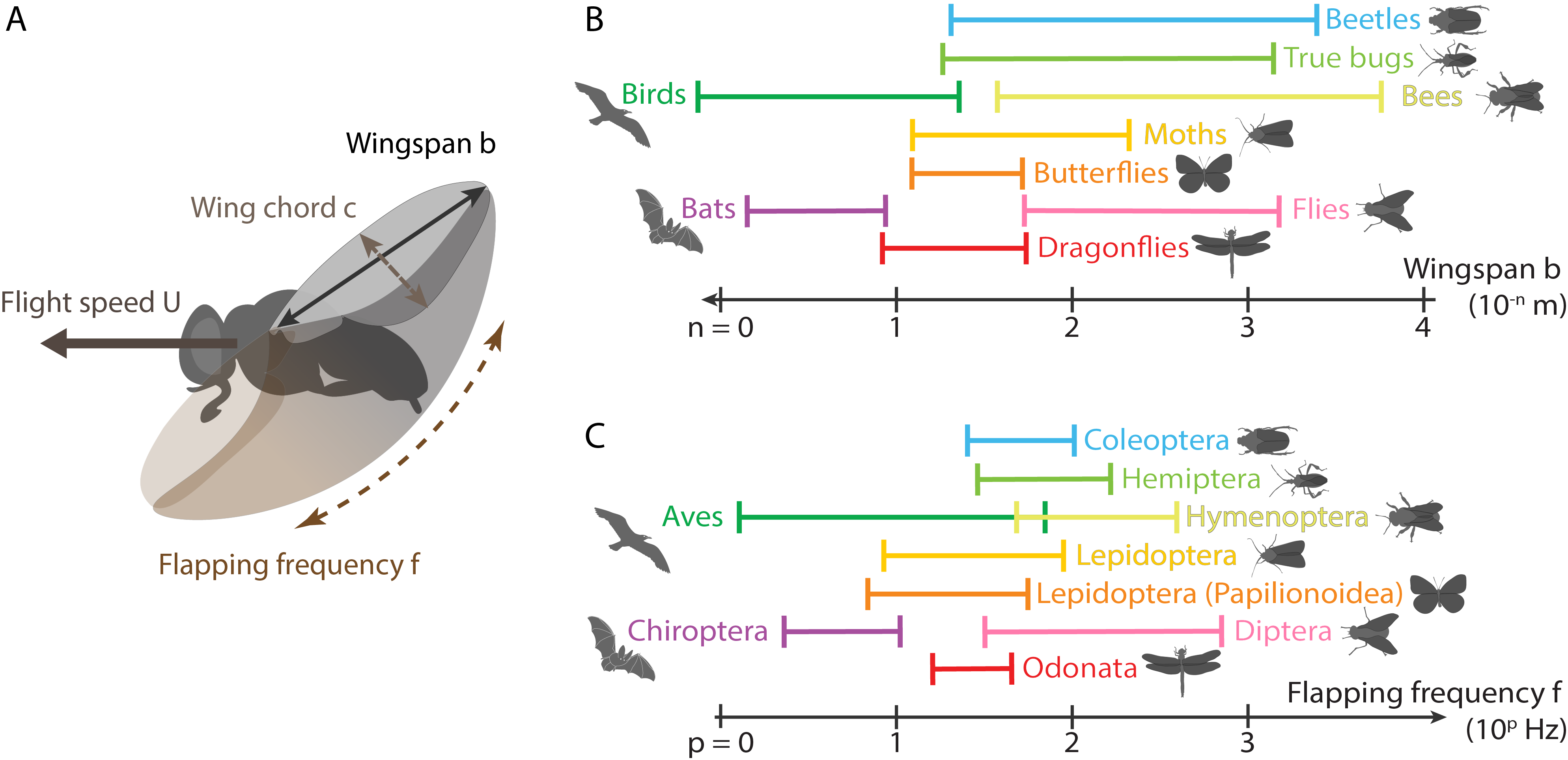}
    \caption{(A) Schematic representation of a flying insect (model: Diptera - \textit{Calliphora vomitoria}). (B-C) Principal groups of flying animal species (vernacular name: B – scientific name: C). Range of wingspan b (B) and flapping frequency f (C) in log scale.}
    \label{fig:diversity}
\end{figure*}

During flapping flight, the wing interacts with the stroke-induced air flow in the atmosphere, sustaining lift and thrust though a wing-flow coupling. Throughout the animal kingdom, a common signature of this coupling is found in the leading-edge vortex formed at the front of the wing during the flapping stroke. The leading-edge vortex is widely recognized as a universal mechanism for lift enhancement in a suite of flying animals, from tiny insects to large birds \cite{Ellington1996, Chin2016, Liu2024}. Observed over a large range of wingspans \cite{Liu2024}, the existence and importance of the leading-edge vortex appear to be independent from the Reynolds number ($\mathrm{Re} = Ub/\nu$, with $U$ the flight speed, $b$ the wingspan of the animal and $\nu$ the kinematic viscosity of the air), with a universal formation time \cite{Sun2025}. This essentially contrasts with aquatic locomotion where the Reynolds number is a key factor in propulsion dynamics \cite{Gazzola2014}. Despite the universality of the leading-edge vortex mechanism, the variety in size, morphology and wingbeat frequency of flying animals makes it cumbersome to construct a unifying framework that explains flight kinematics across animal taxa. Indeed, flying animals span four orders of magnitude in wingspan and three orders of magnitude in flapping frequency (Fig. \ref{fig:diversity}). 

To understand the unifying principles of animal flight as well as how environment plays a part in these principles, we develop a novel turbulence-based scaling framework. Turbulence has gained increasing attention in animal flight research, yet with the focus on only a few model species (bees, hawkmoths, hummingbirds) \cite{Combes2009, OrtegaJimenez2013, Ravi2015, Crall2017, Hejazi2022}. Governed by a multiscale dynamic, turbulence consists in the atmosphere of large-scale vortices up to a few hundred kilometers, like cyclones, down to small-scale vortices of a few hundred micrometers \cite{Nastrom1985, Frisch1995}. Turbulence is thus present over the whole range of flying animal sizes, even to the smallest flyers that have Reynolds numbers on the order of 1, where the surrounding fluid would otherwise be considered laminar when still. Given that the intensity of atmospheric turbulence varies with habitats and climates \cite{Yasuda1988}, animals are expected to have adapted to the intensity of the turbulence they encounter. We thus reexamine animal flight from the perspective of atmospheric turbulence, using a turbulence-based variable to explain animal flight in its ecological context.
\newline

A core variable in turbulence that we can transpose to flying animals is the turbulence energy density spectrum. It represents the density of energy in the air at different wavenumbers $k$, which we will refer to as scales in the rest of this article, associated with vortex size. At first order, we hypothesize that, to avoid getting swept by vortices of its own size and harness the lift from its own leading-edge vortex before it gets dissipated, the minimal energy which the animal has to inject in the atmosphere can be estimated from the turbulence energy density spectrum. The equivalent spectral energy density a flying animal creates is represented by the energy it injects in the atmosphere at its own scale, through its wake. Given the scarcity of such energy measurements, we propose here that the energy injected by the animal into the atmosphere is proportional to the kinetic energy (per unit mass) of the air at the wingtip during the flapping motion $(bf)^2$, where $b$ is the wingspan and $f$ the flapping frequency, which is a function of $b$ depending on the animal.
This energy is concentrated at the scale of the animal’s wingspan, determined by the inverse of the animal wing span $k \sim 1/b$. We write this concentration at the wingspan scale as $\delta(k-1/b)= b\delta(kb-1)$, with $\delta$ being the Dirac function \footnote{For the ease of reading, the Dirac function will be omitted in the notation of $E_{\rm sp}$ in the rest of the manuscript.}. For a primary simplified estimate of the injected energy, we thus define the spectral energy density $E_{\rm sp}(k)$ injected by an animal into the atmosphere by:
\begin{equation}\label{eqn:def}
    E_{\rm sp}(k)= b^3 f^2 = f^2 k^{-3}
\end{equation}
This spectral energy density $E_{\rm sp}$, which we refer to as the animal turbulent energy density, is associated to energy production in the animal wake. It is to be compared with the turbulence spectral energy density in the atmosphere $E_{\rm turb}$, for which the Kolmogorov-Richardson energy cascade is a defining feature \cite{Pope2000}. In the scale range of animals, the energy cascade is expected to follow the inertial scaling $-5/3$ \cite{Frisch1995}, which can be written as:
\begin{equation}\label{eqn:turb}
    E_{\rm turb}(k)=C_K \varepsilon^{2/3} k^{-5/3}
\end{equation}

Where $C_K \sim 0.5$ is the Kolmogorov constant \cite{Sreenivasan1995} and $\varepsilon$ is the atmospheric dissipation rate that quantifies the power of turbulence in the atmosphere. Typical values for $\varepsilon$ at night are $\varepsilon \sim 10^{-4} ~\SI{}{\meter\squared\second^{-3}}$, whereas at daytime it varies between $\varepsilon \sim 10^{-4}~\SI{}{\meter\squared\second^{-3}}$ and $\varepsilon \sim 5 \times 10^{-1}~\SI{}{\meter\squared\second^{-3}}$, depending on the location and altitude \cite{Yasuda1988}.

Combining the animal turbulent energy density for animals across their range of wingspans, we derive scaling relations between the energy density $E_{\rm sp}$ and the scale derived from the wingspan $k=1/b$, in the form $E_{\rm sp}\propto k^\alpha$, with $\alpha$ the exponent to be determined. These scaling relations provide a preliminary estimate of the coefficient that would link $E_{\rm sp}$ and $E_{\rm turb}$. Various scaling relations can be found between the frequency $f$ and the wingspan $b$ in the animal flight literature \cite{WeisFogh1977}. Alternative scaling relations also consider the surface of the wing or the mass of the animal \cite{Sotavalta1947, Jensen2024, Liu2025}. As we focus here on a mass-specific energy, we will concentrate on three scaling relations of $f$ and $b$, which we refer to as rules, established and named by Weis-Fogh \cite{WeisFogh1977}. 

First, the relation known as the ``aerodynamic'' rule stems from a force balance between lift and weight in hovering flight, expressed as $f \propto b^{-1/2}$ \cite{Lighthill1977, WeisFogh1977}. Based on our definition of $E_{\rm sp}$ (Eq. \ref{eqn:def}), this translates into a $-2$ scaling relation ($\alpha_{\rm aero}=-2$) for the energy density with $k \sim1/b$:
\begin{equation}\label{eqn:aero}
    E_{\rm sp}(k)=f^2 k^{-3} \propto (b^{-1/2})^2 k^{-3} \propto k^{-2}
\end{equation}

Second, the “general interspecific’’ rule, expressed as $f\propto b^{-1}$, is observed for well-defined insect families \cite{WeisFogh1977} and can be derived from a skeletal-stress constraint \cite{Lighthill1977}. Flapping flight generates mechanical stress on the wing skeletal structure, bounded above by how the wing materials are resistant to stress and bending. This rule constitutes an upper mechanical limit to the flapping frequency and yields a $-1$ scaling relation with $k \sim 1/b$ ($\alpha_{\rm stress}=-1$) in our context of the animal turbulent energy density: 
\begin{equation}\label{eqn:shear}
    E_{\rm sp}(k)= f^2 k^{-3}\propto(b^{-1})^2 k^{-3}\propto k^{-1}
\end{equation}

Third, an “intraspecific” rule is observed as $f\propto b^{-2/3}$  and supported by a size-independent mass-specific power output \cite{WeisFogh1977}.  As its name implies, the intraspecific rule is primarily encountered when studying individual dispersion within a single species, where the mass-specific power output is determined by the species physiology. Applying this rule to the animal turbulent energy density results in a $-5/3$ scaling relation with $k \sim 1/b$ ($\alpha_{\rm power}=-5/3$): 
\begin{equation}\label{eqn:power}
    E_{\rm sp}(k)=f^2 k^(-3)\propto (b^{-2/3})^2 k^{-3}\propto k^{-5/3}
\end{equation}

Complementary to these three rules from the literature, we can add a fourth scaling prediction based on a physiological limit on the flapping frequency. Animals cannot flap their wings faster than their muscles can elastically contract, meaning that there is a limit $f=f_{\rm max}$ constant and independent of the wingspan b. This physiological constraint translates into a $-3$ scaling relation with $k \sim 1/b$ ($\alpha_{\rm physio}=-3$): 
\begin{equation}\label{eqn:freq}
    E_{\rm sp}(k)=f^2 k^{-3} \propto f_{\rm max}^2 k^{-3} \propto k^{-3} 
\end{equation}

Within these four theoretical scaling laws, the derived $-5/3$ scaling relation establishes the basis for a direct comparison between the atmospheric turbulence spectrum and its animal counterpart. Interestingly, the $-5/3$ relations in turbulence and in animal flight result from similar considerations on power. The locally constant energy dissipation rate $\varepsilon$ in turbulence can be interpreted as a constant mass-specific power output in animal flight. Therefore, similarly to Eq. \ref{eqn:turb} for turbulence, we can write Eq. \ref{eqn:power} in its complete dimensional form, with $C$ being a dimensionless constant and $P_m$ the mass-specific power output:
\begin{equation}\label{eqn:fullpower}
    E_{\rm sp}(k)=C P_m^{2/3} k^{-5/3}
\end{equation}

The $-5/3$ exponent is then constrained by the physical dimensions, unless a fourth parameter is introduced, either a second length or a dependency in the mass-specific power. Determining which predicted exponent is best suited to match experimental data is then key to identify which parameters and physical phenomena govern animal flight allometry.
\newline

\begin{figure*}
    \centering
    \includegraphics[width=0.9\linewidth]{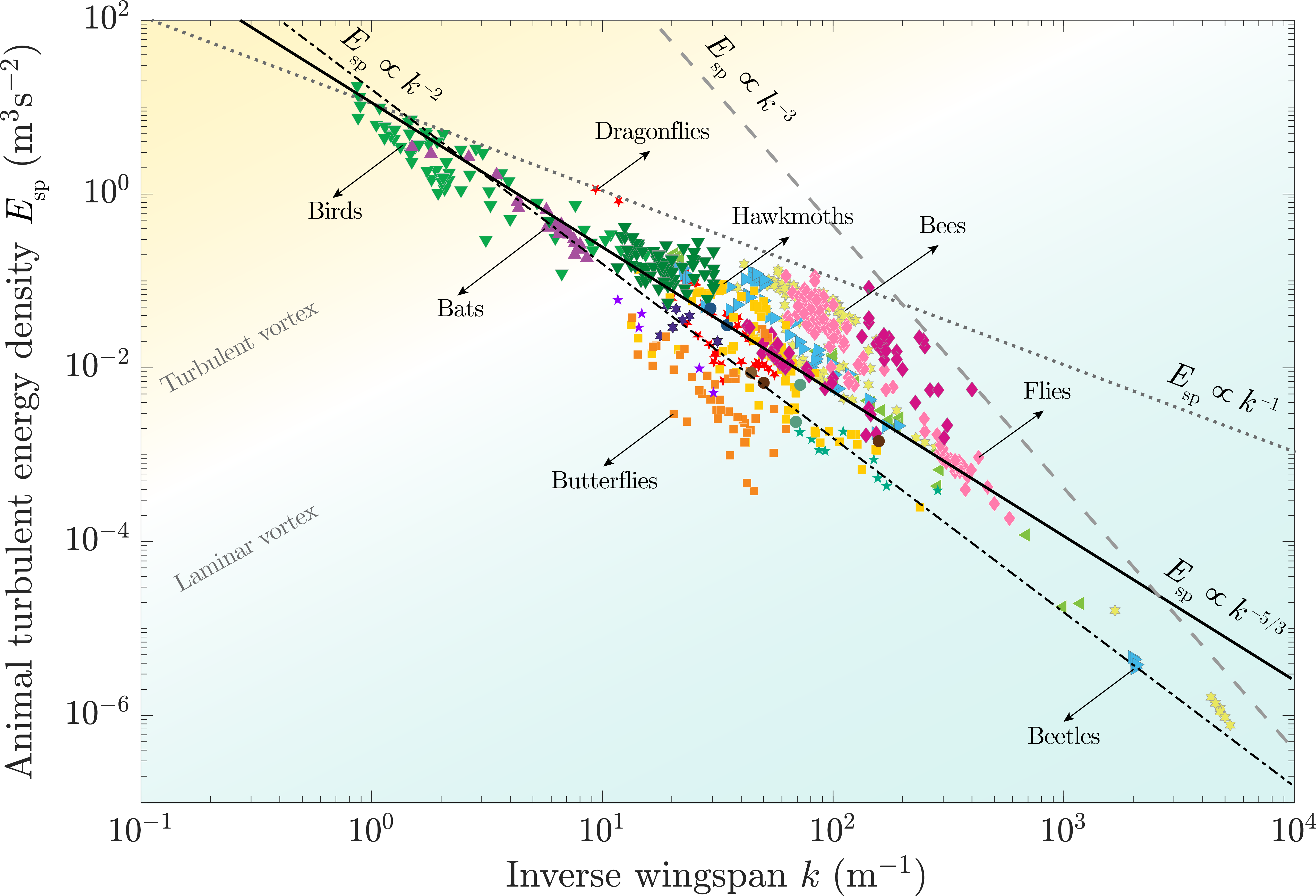}
    \caption{Animal turbulent energy density as a function of the inverse wingspan (in loglog scale). Each point represents an experimental data point from the literature \cite{Arimoto2014,Bullen2002,Cheng2018,Dudley1990,Farisenkov2022,Greenewalt1962,Ha2013,Hefler2020,Hocking1953,Imafuku2024,Kawahara2024,Kutsch1991,Kutsch2002,LeRoy2026,Liu2021,Lyu2025,Magnan1934,Norberg1972,OCallaghan2022,Oehme1959,Osborne1951,Pennycuick1990,Pennycuick2001,Reed1942,Sato1997,Sotavalta1947,Sotavalta1952,Sotavalta1954,Stolpe1939,Stresemann1932,Tercel2018,Williams2025,Yoshida2017}. Vertebrate species are grouped by class except hummingbirds (Trochilidae family). Invertebrate species are grouped by orders except Lepidoptera and Diptera, which are grouped following the previous nomenclature of suborders Heterocera (moths) – Rhopalocera (butterflies) and Nematocera (mosquitoes and craneflies) – Brachycera (flies). Detailed legend available in Fig. \ref{sfig:legend}. Background color represents the flow regime at the wingtip of the animal in still air: turbulent for $\mathrm{Re}_{\rm wingtip}=b^2 f/\nu > 10^4$ (yellow) and laminar for $\mathrm{Re}_{\rm wingtip}=b^2 f/\nu < 10^3$ (blue).}
    \label{fig:scaling}
\end{figure*}

Testing our parametrization and equations against published literature data, we plot in Fig. \ref{fig:scaling} the animal turbulent energy density $E_{\rm sp}$ with respect to the inverse wingspan scale $k=1/b$, for over 700 measurements, across 423 species in 25 orders among birds, bats and insects and compiled from 33 empirical studies \cite{Arimoto2014,Bullen2002,Cheng2018,Dudley1990,Farisenkov2022,Greenewalt1962,Ha2013,Hefler2020,Hocking1953,Imafuku2024,Kawahara2024,Kutsch1991,Kutsch2002,LeRoy2026,Liu2021,Lyu2025,Magnan1934,Norberg1972,OCallaghan2022,Oehme1959,Osborne1951,Pennycuick1990,Pennycuick2001,Reed1942,Sato1997,Sotavalta1947,Sotavalta1952,Sotavalta1954,Stolpe1939,Stresemann1932,Tercel2018,Williams2025,Yoshida2017}. The animals varied in wingspan from \SI{250}{\micro\meter} to \SI{1.4}{\meter}, whereas their flapping frequency varied from \SI{0.5}{\hertz} to \SI{850}{\hertz}. No clear separation is observed between vertebrates (birds and bats) and insects. Insects of size about \SI{1}{\milli\meter} are underrepresented in flight kinematics data, likely due to human observation bias \cite{WeisFogh1973, Cheng2016, Lyu2019, Farisenkov2022, Sun2023}. We see that, despite the limited data on small insects, the data is remarkably clustered over nearly 4 orders of magnitude in wingspan and in good agreement with our four theoretical predictions. 

Among the four theoretical predictions, we observe that the ones associated with body mechanical resistance act as limits to flight kinematics across all animals. Despite the dispersion reaching up to three decades in animal turbulent energy at constant size, this dispersion can be bounded by the two scaling relations corresponding to the physical limits of the animal (Eq. \ref{eqn:shear}, \ref{eqn:freq}). The upper limit of energy per scale follows a $-1$ scaling corresponding to $\alpha_{\rm stress}$ from the interspecific rule (dotted line).  Similarly, the smallest scale per energy limit is determined by the physiological limit of maximal flapping frequency giving $\alpha_{\rm physio}$ associated with a $-3$ power law (dashed line). 

Second, the aerodynamic rule (Eq. \ref{eqn:aero}) accurately predicts the energy density $E_{\rm sp}$ of extreme-sized animals but fails to capture the insect diverse flight (dash-dotted line).  Large birds, bats, feather-winged beetles and parasitoid wasps follow a -2 scaling relation, corresponding to $\alpha_{\rm aero}$. Interestingly, the coefficient of the $-2$ scaling strikingly underestimates the energy density of hovering species like hawkmoths and hoverflies. This is particularly surprising given how the aerodynamic scaling is based on a hovering flight model \cite{Lighthill1977}. Yet, adapting the coefficient to hawkmoths or hoverflies would result in a lesser fitness of the scaling relation at extreme sizes, for which the scaling relation is remarkably fitting. This discrepancy may stem from the fact that our turbulence-based parameter $E_{\rm sp}$ is mass-specific, whereas the aerodynamic rule resulting in $\alpha_{\rm aero}$ is drawn from a force balance involving the animal weight.

Constant mass-specific power output appears to primarily predict the quantity of energy $E_{\rm sp}$ injected by the animal into the atmosphere at its own scale. The $-5/3$ scaling relation (solid line) provides the best fit to the experimental data ($R^2=0.7572$). This scaling relation of exponent $\alpha_{\rm power}$ provides a unified scaling across widely different animals in terms of morphology, down to approximately 1 mm wingspan. Though it could be seen as a downgrade to the aerodynamic rule, the resulting dispersion from the best fit is almost symmetrical among insects, with butterflies and bees at the extremes of the dispersion. This symmetry indicates less bias on fitting with $\alpha_{\rm power}$, capturing better the overall trend for $E_{\rm sp}(k)$. If we assume the dimensionless coefficient C in Eq. \ref{eqn:fullpower} to be equal to 1, we can extract the mass-specific power output from the best fit: $P_m=67 \pm \SI{50}{\watt\per\kilo\gram}$. The value range we obtained for $P_m$ is in excellent agreement with the range of mass-specific power output recorded in the literature \cite{Ellington1991, Dudley2000}, despite the large uncertainty coming from the logarithmic fit and the simplifications made in our model.

\begin{figure*}
    \centering
    \includegraphics[width=0.9\linewidth]{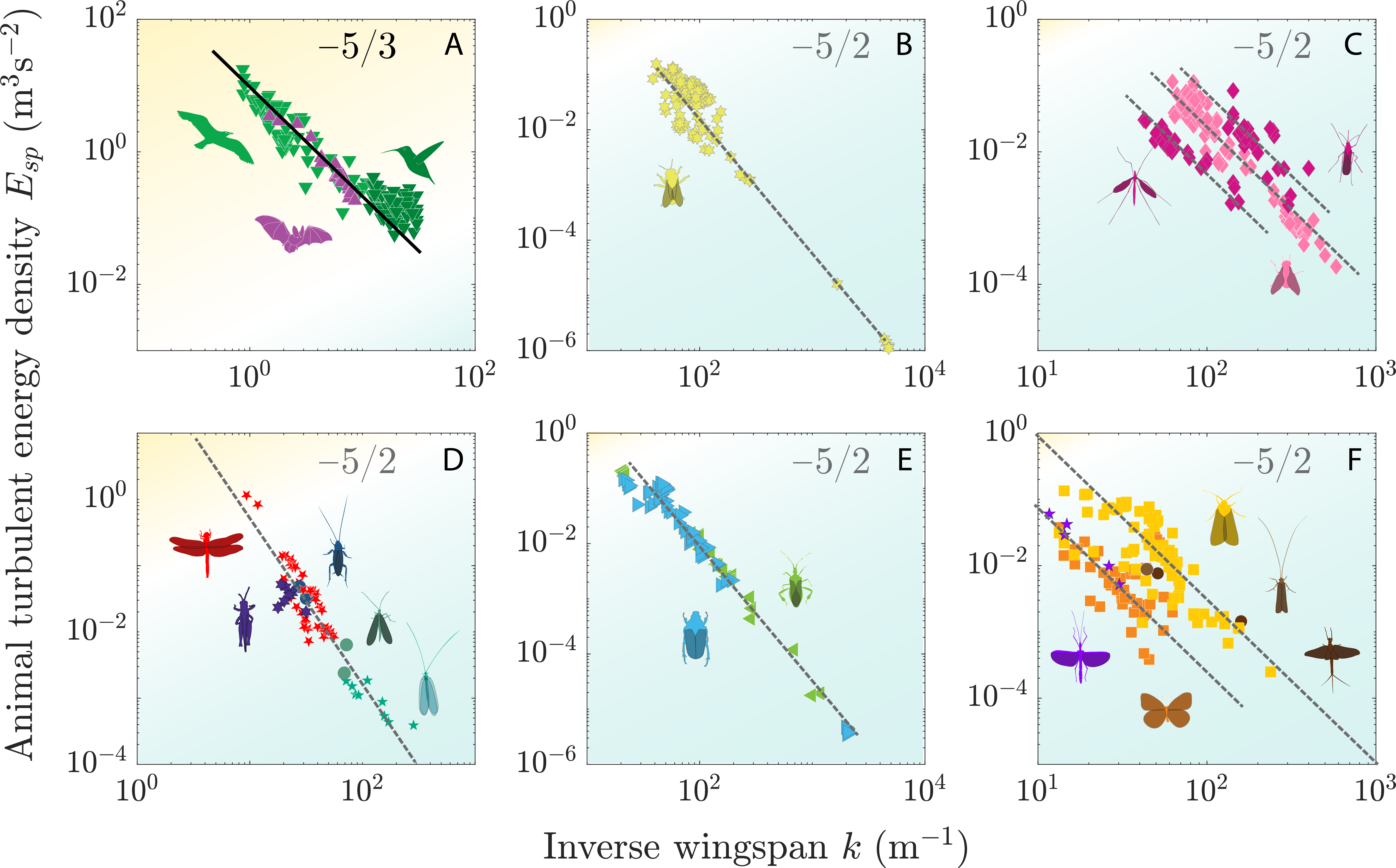}
    \caption{Animal turbulent energy density as a function of the inverse wingspan (in loglog scale) zoomed on different groups of similar wing use (same data as Fig. 2). Vertebrates (A) present a $-5/3$ scaling law as observed for the whole animal kingdom (Fig. \ref{fig:scaling}). Invertebrates (B-F) follow however a $-5/2$ scaling relation when segregated per order, with a coefficient depending on the order. Diptera (C) and Lepidoptera (F) are the only orders for which the coefficient depends on the suborders. Background color represents the flow regime at the wingtip of the animal in still air: turbulent for $\mathrm{Re}_{\rm wingtip}=b^2 f/\nu > 10^4$ (yellow) and laminar for $\mathrm{Re}_{\rm wingtip}=b^2 f/\nu < 10^3$ (blue). Legend and zoom details in Fig. \ref{sfig:legend}.}
    \label{fig:phylogeny}
\end{figure*}

The dispersion of the insect data suggests that the scaling relations currently available still lack a key parameter to capture a physical universality behind the diversity of insect flight. In particular, the fact that the constant mass-specific power output relation (Eq. \ref{eqn:fullpower}) better fits the experimental data considered as a whole compared to the aerodynamic relation raises the question of whether mass-based models are sufficient to describe animal flight \cite{Lighthill1977, Jensen2024}. The choice of our turbulence-inspired variable $E_{\rm sp}$ emphasizes the minute differences in flight kinematics across different insects of similar sizes that are not captured by the hovering model \cite{Lighthill1977}.
\vspace{0.2cm}

Grouping insect species by wing use in flight reveals a novel scaling relation not predicted by the current knowledge of animal flight (Fig. \ref{fig:phylogeny}). In addition to the separation between vertebrates (Fig. \ref{fig:phylogeny}.A) and invertebrates (Fig. \ref{fig:phylogeny}.B-F), we identify eight groups of insect species with closely matching wing use (see Appendix). We here consider wing use as a combination of wing configuration in flight, flapping amplitude and muscle actuation type. Dividing by wing use was motivated by recent works highlighting similarities and differences across the vast diversity of insect flight \cite{LeRoy2025, LeRoy2026, Gau2023}. This division in eight groups, primarily linked with phylogeny, also encompasses behavioral and physiological differences. As shown in Fig. \ref{fig:phylogeny}.B-F, for each of the eight groups, we observe the following scaling relation \footnote{Model tested for best fit significance against the -2 scaling relation using Vuong’s closeness test: $p<0.0001$.}:
\begin{equation}\label{eqn:neo}
    E_{\rm sp}(k)\propto k^{-5/2}
\end{equation}

This novel scaling presents a group-dependent coefficient, linked to wing use in flight. The universality of the scaling exponent suggests a yet common mechanism throughout insect phylogeny. Based on how we combined the species into eight groups of similar wing use and morphology, we expect wing and flight muscle mechanical properties to be determining in shaping the dispersion we observe in Fig. \ref{fig:scaling}. The distinction between the $-5/3$ and the $-5/2$ scaling relations not only separates the insects from the vertebrates, but also corresponds to a separation in the wingtip Reynolds number $\mathrm{Re}_{\rm wingtip}=b^2 f/\nu$ (background colors in Fig. \ref{fig:scaling} and \ref{fig:phylogeny}). When considering the air surrounding the wingtip independently from its environmental context, $\mathrm{Re}_{\rm wingtip}$ assesses the flow nature of the leading-edge vortex. A high wingtip Reynolds number $\mathrm{Re}_{\rm wingtip}>10^4$ corresponds to a spontaneously-turbulent vortex (yellow background), whereas a low wingtip Reynolds number $\mathrm{Re}_{\rm wingtip}<10^3$ is associated to a laminar vortex (blue background), as observed behind forward-swept wings at low Reynolds numbers \cite{Zhang2022}. A similar separation between low and high Reynolds number regimes has been described and modeled in aquatic locomotion \cite{Gazzola2014}, but this contrasts with the observation of universal mechanisms in animal flight \cite{Chin2016, Liu2024, Sun2025}. Observing conjointly universal and disparate phenomena across flying animals therefore suggests that animal flight physics is inherently dependent on the environment in which the animal flies, as it cannot be unified by focusing solely on the animal itself.

By extending the scaling approach of animal locomotion to the energetics of the surrounding air through a turbulence-based parametrization, we identify a unifying scaling relation in insect flight. This scaling relation with a unique exponent of $-5/2$ suggests that, despite the vast morphological diversity of insects, their flight kinematics can be approximated with a universal model. Our turbulence-based parametrization also provides intriguing insights about animal flight across animal classes. Interestingly, the $-5/3$ scaling relation of vertebrates not only fits a large amount of insect species but also matches the inertial scaling of turbulence \cite{Frisch1995, Pope2000}, which questions the influence of atmospheric properties on the selection of flight kinematics and insect flight performance \cite{Engels2019}. Simultaneous recordings of atmospheric properties and flight kinematics across species in an ecologically-relevant context will then prove key to our understanding of animal flight. 
\\

\begin{acknowledgments}

This publication is part of project ``On the Fly: understanding multiscale turbulence through animal flight'' (VI.Veni.232.249) funded by the Dutch Research Council (NWO) as part of the Talent Programme Veni Science domain and awarded to AG. CJvdK acknowledges funding from Human Frontiers in Science Program (RGP023/2023, DOI: 10.52044/HFSP.RGP0232023.pc.gr.168611) and AFOSR (FA8655‐23‐1‐7049).
We thank D. G. Stavenga for helping structure this publication, and we are grateful to M. Obligado, A.-J. Buchner and M. Bourgoin for pre-reviewing this work. \end{acknowledgments}

\appendix

\section{Appendix}\label{sec:appendix}

\subsection{Description of the choice for grouping insect families in Fig. \ref{fig:phylogeny}}
First (Fig. \ref{fig:phylogeny}.B), we consider the bees and wasps (Hymenoptera) which constitute a large order with little wing shape differences: forewings and hindwings getting hooked to one another in flight. We separated (Fig \ref{fig:phylogeny}.C) flies (Brachycera) from craneflies (Tipulidae and Limoniidae) on the lower side, and, on the higher side, the remaining Diptera families among which are mosquitoes (Culicidae, Chironomidae and Bibionidae). We distinguish three groups among flies (Diptera) based on flight behavior, sharing the two-wing arrangement. Such segregation of Diptera species was motivated by the recent extensive studies on Diptera flight \cite{LeRoy2026}. Linking to muscle actuation, we group insects which four wings are actuated by synchronous muscles and not overlapping during flight (Fig. \ref{fig:phylogeny}.D):  dragonflies (Odonata), scorpionflies (Mecoptera), lacewings (Neuroptera) and grasshoppers (Orthoptera) and cockroaches (Blattodea). Another group (Fig. \ref{fig:phylogeny}.E) is found in true bugs (Hemiptera) and beetles (Coleoptera) which possess independent forewings and hindwings actuated by asynchronous muscles. Finally (Fig. \ref{fig:phylogeny}.F), we grouped butterflies (Papilionoidea) with stick insects (Phasmatodea) based on their low wing aspect ratio, and separated them from moths (non-butterfly Lepidoptera) grouped with caddisflies (Trichoptera) and mayflies (Ephemeroptera). This last grouping is motivated by the common wing use of overlapping wings actuated by synchronous indirect flight muscles.

\begin{figure*}
    \centering
    \includegraphics[width=0.85\linewidth]{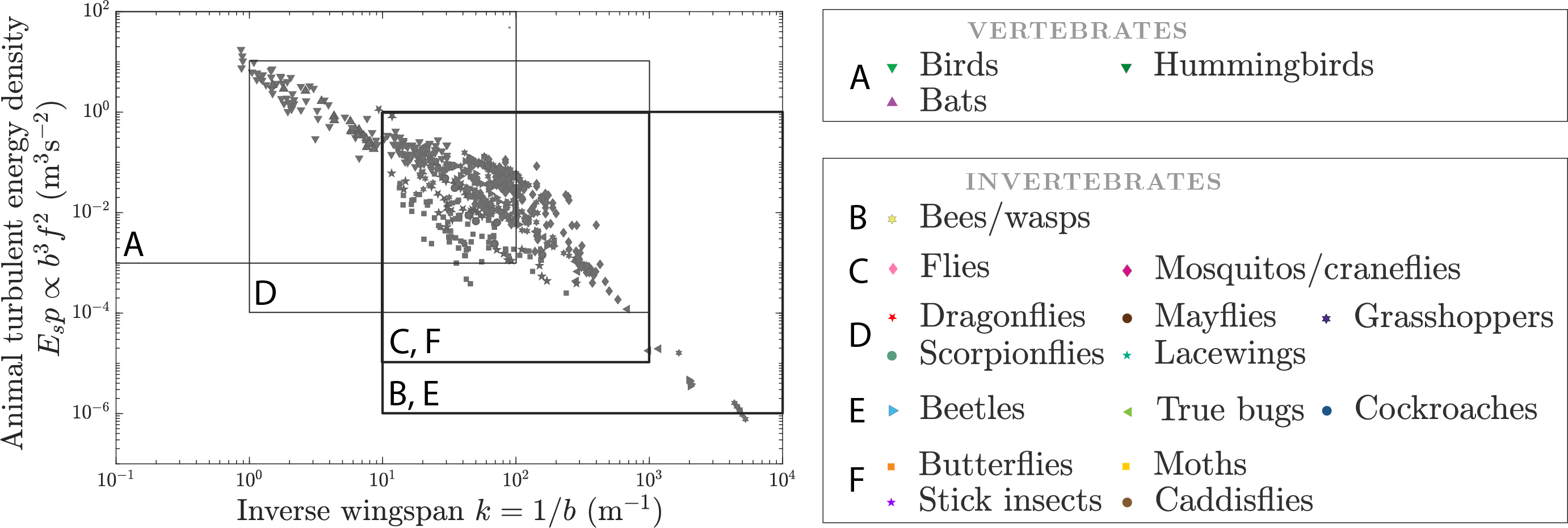}
    \caption{Left: zoom details of Fig. \ref{fig:phylogeny}. Right: detailed legend common to Fig. \ref{fig:scaling} and Fig. \ref{fig:phylogeny}.}
    \label{sfig:legend}
\end{figure*}

 \begin{table*}
 \caption{\label{stab:suptable} Values of $R^2$ and $C$ for the best fit $\log_{10} E_{\rm sp} = -\alpha \log_{10} k + C$ with different exponents $\alpha$, obtained using MATLAB$^\circledR$ Curve Fitting toolbox. Highlighted in bold are the highest values within $5\%$}
 \begin{ruledtabular}
 \begin{tabular}{c c cccccc}
 
 Phylum & Group  & $R^2$ $\alpha_{\rm power}$ & $R^2$ $\alpha = -5/2$ & $R^2$ $\alpha_{\rm aero}$ & Best $\alpha$ & $R^2$ best fit & Coefficient $C$\footnotemark[1]\\
\\[-0.5em]
\hline
\\[-0.5em]
\multirow{16}{*}{Arthropoda} & Cranefly morphs\footnotemark[2] &	\textbf{0.8978} & 0.8214 & \textbf{0.9092} & -1.902 & 0.9117 & 2.619\\
& Mosquito morphs\footnotemark[2] &	0.5396 & \textbf{0.5846} & \textbf{0.5756} & -2.334 & 0.5876 & 3.866\\
& Brachycera & 0.8011 & \textbf{0.9488} & 0.8804 & -2.814 & 0.9607 & 3.366\\
& Diptera (all) &	0.66 & 0.6514 & \textbf{0.6836} & -2.052 & 0.6841 & 3.322\\
\\[-0.5em]

& Butterflies\footnotemark[3] &	\textit{0.24} & \textit{0.09} & \textit{0.20} & \textit{-1.39} &	\textit{0.25} &	– \\
& Moths &	\textbf{0.61} &	0.58 & \textbf{0.62} & -2.00 & 0.62 & 2.372\\
& Butterfly morphs\footnotemark[4] & 0.5918 & \textbf{0.6481} & 0.6335 & -2.377 & 0.6498 & 1.252\\
& Moth morphs\footnotemark[4] &	0.6667 & \textbf{0.7431} & 0.7178 & -2.455 & 0.7433 & 2.422\\
& Lepidoptera (all)\footnotemark[3] & \textit{0.2819} & \textit{0.1595} & \textit{0.254} & \textit{-1.502} & \textit{0.2853} & – \\
\\[-0.5em]
& Odonata	& 0.6514 & \textbf{0.7648} & 0.7137 & -2.75 &	0.7712	& 2.25\\
\\[-0.5em]
& Coleoptera &	0.8654 & \textbf{0.9586} & 0.9299 & -2.426 & 0.9595 & 2.915\\
\\[-0.5em]
& Hemiptera &	0.9128 & \textbf{0.9688} & \textbf{0.9672} & -2.258 & 0.98 & 2.92\\
\\[-0.5em]
& Hymenoptera &	0.8268 & \textbf{0.9296} & 0.8928 & -2.497 & 0.9296 & 3.342\\
\\[-0.5em]
\hline
\\[-0.5em]
\multirow{4}{*}{Chordata} & Aves &	\textbf{0.7919}	& 0.3389	& 0.6786 & -1.416 & 0.8174	& 0.9379 (-5/3)\\
& Hummingbirds\footnotemark[3] & \textit{-0.0608} & \textit{-0.9778} & \textit{-0.356} & – & – & – \\
\\[-0.5em]
& Chiroptera & \textbf{0.9424} & 0.7731	& 0.9259 & -1.753 & 0.9447 & 0.9362 (-5/3)\\
\\[-0.5em]
\hline
\\[-0.5em]
& All Orders & \textbf{0.7572} & 0.4573	& 0.6915 & -1.531 & 0.7632 & 1.239 (-5/3) \\

 \end{tabular}
 \end{ruledtabular}
 
% \footnotetext[1]{Here's the first, from Ref.~\onlinecite{feyn54}.}
 \footnotetext[1]{Coefficient $C$ obtained with fitting $\log_{10} E_{\rm sp} = -5/2 \log_{10} k + C$, unless stated otherwise in parenthesis}
 \footnotetext[2]{We separate the former infraorder Nematocera into cranefly morphs (Tipulidae, Limoniidae) and mosquito morphs (Culicidae, Chironomiidae)}
 \footnotetext[3]{For these groups, best fit $R^2$ for a power law was below 0.5.}
 \footnotetext[4]{We group Lepidoptera between butterfly morphs and moth morphs, to accommodate for their wing shape. In particular, Hesperiidae are considered moth morphs whereas Saturniidae are butterfly morphs.}
% \footnotetext[5]{And etc.}
\end{table*}


\bibliography{bib_260604}% Produces the bibliography via BibTeX.

\end{document}